\documentclass[12pt,twoside]{article}   
\usepackage[super,sort,comma]{natbib}

\usepackage{amsmath}
\usepackage{fancyhdr}		
\usepackage{booktabs}
\usepackage{multirow}
\usepackage{xcolor}
\usepackage{colortbl}
\definecolor{revised}{rgb}{0,0,0}
\definecolor{question}{rgb}{0,0,0}

\usepackage[section]{placeins}   %

\usepackage{graphicx}

\makeatletter \renewcommand\@biblabel[1]{$^{#1}$} \makeatother
\newcommand{\cen}[1]{\begin{center} #1 \end{center}}

\usepackage[mathlines]{lineno}

\usepackage{subcaption}
\usepackage{hyperref}
\usepackage{adjustbox}
\usepackage{amsmath}
\usepackage{amssymb}
\usepackage{graphics}
\usepackage{colortbl}
\hypersetup{ colorlinks,
    citecolor=blue,
    filecolor=blue,
    linkcolor=blue,
    urlcolor=blue
}

\usepackage{xcolor}

\definecolor{gray}{rgb}{0.6,0.6,0.6}
\definecolor{red}{rgb}{0.85,0,0}
\definecolor{green}{rgb}{0,0.85,0}
\definecolor{blue}{rgb}{0,0,0.85}
\definecolor{beige}{rgb}{0.92,0.87,0.78}
\usepackage[all]{hypcap}    
\usepackage{url}

\begin{document}

\cen{\sf {\Large {\bfseries {On the Degrees of Freedom of Gridded Control Points in Learning-Based Medical Image Registration}} \\  
\vspace*{10mm}
Wen Yan$^1$, Qianye Yang$^1$, Yipei Wang$^1$, Shonit Punwani$^2$, Mark Emberton$^3$, Vasilis Stavrinides$^{4,5}$, Yipeng Hu$^1$, Dean Barratt$^1$ }\\
$^1$ UCL Hawkes Institute; Department of Medical Physics and Biomedical Engineering, University College London, Gower St, London WC1E 6BT, London, U.K.\\
$^2$ Centre for Medical Imaging, Division of Medicine, University College London,  Gower St, London WC1E 6BT, London, U.K.\\
$^3$ Division of Surgery and Interventional Science, University College London, 10 Pond St, London NW3 2PS, U.K.\\
$^4$ Cancer Institute, Urology Department, UCL Hospital, University College London, 16-18 Westmoreland St, London W1G 8PH, U.K\\
$^5$ Radiology Department, Imperial College Healthcare,  The Bays, S Wharf Rd, London W2 1NY, U.K.
}
\pagenumbering{roman}
\setcounter{page}{1}
\pagestyle{plain}
corr email: wen-yan@ucl.ac.uk \\


\begin{abstract}
\noindent{\bf Background:} Many registration problems are ill-posed in homogeneous/noisy regions, and dense voxel-wise decoders can be unnecessarily high-dimensional. A sparse control-point parameterisation provides a compact, smooth deformation representation while reducing memory and improving stability.\\
{\bf Purpose:} This work investigates the required control points for learning-based registration network development. In particular, as sparse as $5\times5\times5$ control points are configured and compared with alternative approaches, including those using scattered control points and displacements sampled at every voxel, i.e. dense displacement fields. \\
{\bf Method:} We present GridReg, a learning-based registration framework that
replaces dense voxel-wise decoding with displacement predictions at a sparse
grid of control points. This design substantially cuts the parameter count and
memory while retaining registration accuracy. Multiscale 3D encoder feature
maps are flattened into a 1D token sequence with positional encoding to retain
spatial context. The model then predicts a sparse gridded deformation field
using a cross-attention module: Each control point attends to encoder tokens
within its local grid neighborhood to estimate its displacement, which is sub-sequently interpolated to a dense field. We further introduce grid-adaptive
training, enabling an adaptive model to operate at multiple grid sizes at inference
without retraining.
\\
{\bf Results: }
This work quantitatively demonstrates the benefits of using sparse
grids. Using three data sets for registering prostate gland, pelvic organs
and neurological structures, the experimental results suggest a much
improved computational efficiency, due to the prediction of sparse-grid-sampled displacements. Alternatively, the superior registration performance
was obtained using the proposed approach, with the similiar or less compute cost, compared with existing algorithms that predict DDFs (e.g., Vox-
elMorph/TransMorph) or displacements sampled on scattered key points
(KeyMorph). \\
{\bf Conclusion:} We conclude that predicting sparsely gridded displacements pro-
vides reduced computational cost and/or improved performance, independent
of the encoder architecture, and can be readily implemented. Therefore,
GridReg should potentially be considered for many registration tasks with adaptive grid sizes. 
The code is available via \url{git@github.com:yanwenCi/GridReg.git}.

\end{abstract}




\setlength{\baselineskip}{0.7cm}      

\pagenumbering{arabic}
\pagestyle{fancy}

\section{Introduction}
{M}{edical} image registration aligns images from different modalities, patients, or time points to the same spatial coordinates, with many clinical utilities~\citep{hill2001medical}. 
Traditional pairwise image registration methods optimise parametric or non-parametric transformation models iteratively given a pair of images~\citep{fitzpatrick2000image}. 
Learning-based image registration enables a data-driven approach with fast inference to predict transformations that represent spatial correspondences~\citep{balakrishnan2019voxelmorph, chen2022transmorph, evan2022keymorph, hoffmann2021synthmorph, chen2024survey}. Examples are discussed in Sec.~\ref{intro:literature}.

\begin{figure}[htb]
    \centering
    \includegraphics[width=0.8\linewidth]{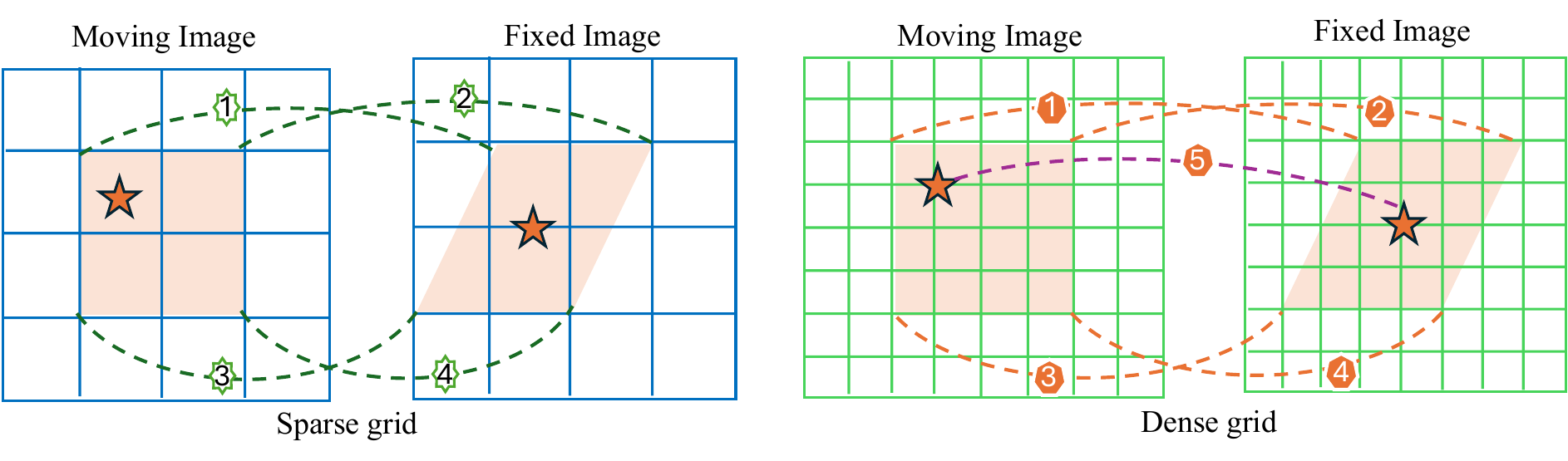}
        \caption{Demonstration of an illustrative example of false-positive correspondence (orange stars), possibly caused by similar intensity patterns (e.g., unsupervised loss similarity) or label errors (e.g., segmentation noise). A sparse low-resolution grid helps filter out such a noise, as stronger ``true'' correspondences (four corners of the shaded ROIs) dominate. In contrast, a finer high-resolution grid may amplify the noise, distorting additional control points within the ROIs and leading to inaccurate registration. }
        \label{fig:grid}
\end{figure}

An interesting observation in the development of learning-based image registration techniques is that early deep learning approaches largely bypassed the use of gridded control points for modeling deformation~\citep{balakrishnan2019voxelmorph, chen2022transmorph, evan2022keymorph, hoffmann2021synthmorph, 9815265}. Instead, these methods commonly predicted dense displacement fields (DDFs), in which a displacement vector is estimated at every voxel location. This architectural choice is likely influenced by the evolution of computing capabilities, particularly the advent of high-performance GPUs and parallelised training frameworks, which have substantially reduced the computational burden associated with predicting and optimising per-voxel displacements. 
Dense displacement fields can be viewed as an extreme case of grid-based models where the grid resolution equals the image resolution — effectively treating every voxel as a control point.

A higher-resolution control grid, defined by a larger number of sampled control points, indeed allows the model to represent more localised, fine-scale transformations. However, this increase in resolution does not necessarily improve registration performance~\cite{elastixManual520, wang2019review}. When the transform has excessively high DoF in medical imaging and similar domains, many regions of interest — such as homogeneous tissue areas or regions with strong random noise — inherently lack well-defined local correspondences. Allowing the network to resolve extremely localised deformations under dense transformation may be unnecessary or even detrimental. It can introduce noise, overfitting, or anatomically implausible correspondences, especially if the similarity metrics used during training fail to distinguish meaningful local variation from spurious noise.
Moreover, increasing the grid density comes at a substantial computational cost. This burden is particularly observed in the decoder, which is responsible for reconstructing latent representations into high-resolution structured deformation outputs. Such decoders typically rely on multi-scale feature integration and upsampling mechanisms, which become progressively more complex and expensive as the output resolution increases~\citep{leng2013medical}. Coarser control grids reduce the degrees of freedom, acting as an implicit regulariser that limits high-frequency, voxel-level warps and can reduce sensitivity to unreliable local correspondences.
As illustrated in Figure~\ref{fig:grid}, the choice of grid resolution plays a crucial role in controlling the granularity of the estimated transformation field. A coarser grid may act as a form of implicit denoising or regularisation, effectively filtering out high-frequency similarity that is not consistently supported by the underlying anatomy or image evidence. This insight suggests that, depending on the application context and anatomical region, using a coarser control grid can lead to more anatomically meaningful and computationally efficient deformation estimates.

\textcolor{revised}{
Our approach differs from pyramidal registration, in which downsampled images still yield a high-DoF dense field via a learned high-resolution decoder~\cite{balakrishnan2019voxelmorph, dalca2019unsupervised}. Instead, we predict a sparse control-point displacement field from flattened 1D feature tokens with positional encoding, and then reconstruct a dense voxel-wise displacement field via interpolation (e.g., trilinear or B-spline~\cite{tustison2013explicit}). Constraining DoF of the predicted transformation acts as built-in spatial regularisation, reducing overfitting in noisy or homogeneous regions. Crucially, low DoF is not low expressivity, where spline interpolation can produce smooth yet highly non-linear, non-rigid deformations. We do not claim that high-resolution displacement fields are never useful; rather, their necessity is often overestimated.}
\textcolor{revised}{
 Our main contributions are:
\begin{itemize}
    \item We introduce a lightweight decoder that projects multi-scale 3D features into control-point embeddings and predicts sparse displacements via local cross-attention.
    \item We propose a grid-adaptive framework that trains across multiple control-grid resolutions and selects an effective grid density for each dataset via validation-driven model selection.
    \item We achieve comparable or improved registration accuracy relative to voxel-wise DDF decoders, while substantially reducing memory and computation and improving deformation regularity, which is attractive for computation-limited settings.
\end{itemize}
}

\section{Related Works}
	\label{intro:literature}
The degree-of-freedom (DoF) of gridded control points reflects the resolution of correspondence in registration, discussed above. A higher DoF (dense) implies the ability to capture finer, local correspondence, while a lower DoF represents coarser correspondence. For scattered control points, correspondence resolution varies spatially, depending on its local sampling mechanisms for allowed DoF per unit space (volume in 3D). The spatial adaptability from scattered points allows versatility and, arguably, efficiency, but requires optimisation or predefined with prior knowledge. In non-learning-based algorithms based on iterative optimisation, these two categories of transformation representation are closely related to free-form deformation~\citep{796284} and point-feature-based method~\citep{AUDETTE2000201}. We discuss example studies in the context of learning-based algorithms as follows. 

\subsection{Gridded control points}

Spatial Transformation Network (STN)~\citep{jaderberg2015spatial} applies a predicted dense displacement field through a differentiable sampling operator, enabling end-to-end optimisation using image-domain losses between the warped source and the target~\citep{shan2018unsupervisedendtoendlearningdeformable}. 
Based on this, much work using encoder-decoder structured networks predicts the transfoThese works utilised unsupervised methods to predict the displacement field between two real-world images by minimising image similarity loss between the warped source image and the target image.e works utilized unsupervised methods to predict the displacement filed between two real-world images by minimizing image similarity loss between the warped source image and the target image. In the case of complex tasks like multi-modality registration, refining the dense deformation field (DDF) requires a more focused investigation into training strategies. Yang proposed a work~\citep{9815265} that benefits from exploiting additional images during training for cross-modalities registration. 
Zhou~\citep{zhou2024differentiable} proposed a structured cross-modality latent space to represent 2D-pixel features and 3D features via a differentiable probabilistic Perspective-n-Points solver. Hussain~\citep{hussain2025hierarchical} proposes a hierarchical cross-attention-based transformer structure to extract features of fixed and moving images at multiple scales, enabling more precise alignment and improved registration accuracy. This structure strengthens the ability of decoder, enabling it to facilitate progressive deformation field refinement across multiple resolution levels.

\subsection{Scattered control points}
The scattered control points method can utilise manually selected anatomical landmarks, aiming to register specific regions of interest~\citep{pantazis2010comparison, lange20093d}. 
Otherwise, it can automatically detect key points using feature extraction methods.  
Keymorph~\citep{evan2022keymorph} utilised a centre of mass layer to generate corresponding key points for source and target images, respectively, and produced a deformation field based on these paired key points. Wang~\citep{wang2024brainmorph} extended KeyMorph as a tool that supports multi-modal, pairwise, and scalable groupwise registration, solving 3D rigid, affine, and nonlinear registration in brain registration.
Ekvall et al.~\citep{ekvall2024spatial} presented an unsupervised spatial landmark detection and registration network to address the challenges in nonlinear deformations between histological tissue sections. Zachary et al.~\citep{baum2021real} present a Free Point Transformer (FPT) in which the scattered points are sampled to represent the surface of the organ via a separate segmentation process for non-rigid point-set registration. Fu et al.~\citep{fu2021biomechanically} generate volumetric prostate point clouds from segmented prostate masks using tetrahedron meshing and then use a point cloud matching network to obtain the deformation field for registration. 
SuperPoint~\citep{detone2018superpoint} operates on full-size images and jointly computes pixel-level interest point locations and associated descriptors in one forward pass using homographic adaptation. 
Zhang et al.~\citep{9658276} proposed a two-stage method for multi-model image registration. In the coarse stage, the method uses a key-point based method inherited from~\citep{detone2018superpoint} to generate a course deformation field from the segmentation map, and in the fine stage, the method uses a voxel-based method to refine the deformation field.

\section{Methods}

\subsection{Predicting gridded displacement fields}\label{sec.method-A}
Let 
\(
\mathcal{I}=\{(\mathbf{F}_j,\mathbf{M}_j)\mid j=1,\dots,J\}
\)
be a dataset of $J$ source–target image pairs, where each image is a 3D tensor 
\(\mathbf{F}_j,\mathbf{M}_j \in \mathbb{R}^{W\times H\times D}\).
Define the voxel domain 
\(\Omega=\{1,\dots,W\}\times\{1,\dots,H\}\times\{1,\dots,D\}\)
and \(N=|\Omega|=WHD\).
We write the vectorized forms as
\(
\mathbf{f}_j \in\mathbb{R}^{N},
\mathbf{m}_j \in\mathbb{R}^{N},
\)
so that
\(
\mathbf{f}_j=[\,f_j(\mathbf{x}_1),\dots,f_j(\mathbf{x}_N)\,]^\top
\)
with voxel coordinates 
\(\mathbf{x}_n=(x_n,y_n,z_n)\in\Omega\).
A spatial transform is represented by a dense displacement field
\(\mathbf{T}_j:\Omega\!\to\!\mathbb{R}^3\) over the voxel domain
\(\Omega=\{1{:}W\}\times\{1{:}H\}\times\{1{:}D\}\).
We denote warping by \(\mathcal{W}(\mathbf{M}_j,\mathbf{T}_j)=\mathbf{M}_j\circ\mathbf{T}_j\), where $\circ$ denotes spatial transform operation~\citep{jaderberg2015spatial}. 
Learning-based image registration optimises the model weights to find transformations that align each pair of source and target 
images. This can be formulated as an optimisation problem: 
\(
\arg\min_{\mathbf{T}_j}\sum_{n=0}^N \mathcal{L}(\mathbf{M}_j \circ \mathbf{T}_j, \mathbf{F}_j ) 
\label{eq:object1}
\)
, where and $\mathcal{L}$ is a registration loss function of the transformed moving image and the fixed image.
\textcolor{revised}{In this study, we propose a registration method based on a sparse gridded displacement field (GDF) with size $G = g_w \times g_h \times g_d$. 
}
\begin{figure*}[htb]   
     \centering
    \includegraphics[width=\linewidth]{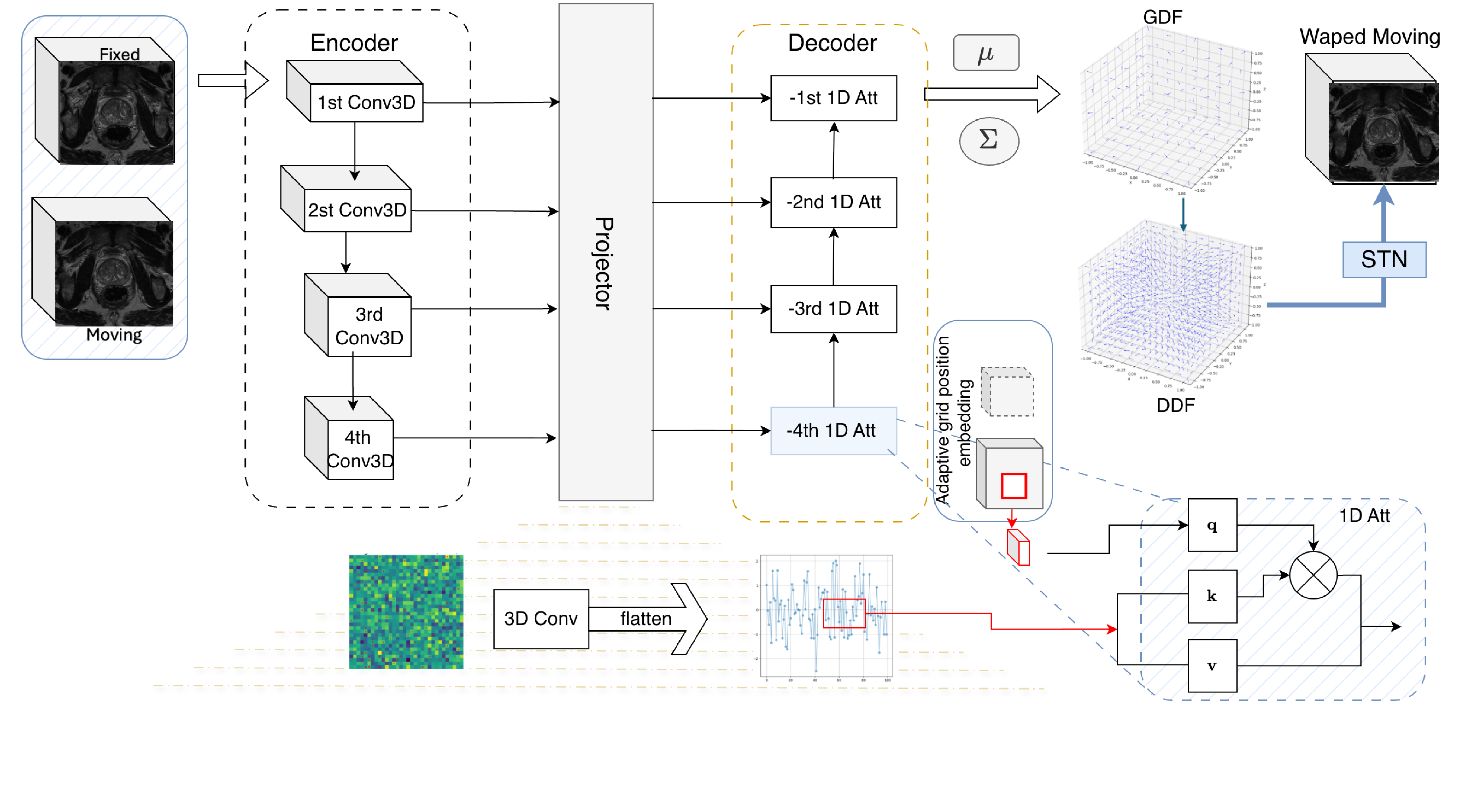}
    \caption{\textcolor{revised}{Illustration of GridReg. The encoder extracts multi-scale 3D feature maps; at each scale, a skip-projection flattens features into 1D tokens that are fused with grid-cell queries via (local) attention. The resulting features are mapped to a sparse control-point displacement field via Bayesian integration, and a dense displacement field (DDF) is obtained by interpolation (e.g., trilinear, B-spline, or transposed convolution).}}
    \label{fig:intro-net}
\end{figure*}
\textcolor{revised}{As illustrated in Figure~\ref{fig:intro-net}, we describe a simple, easy-to-implement registration network, GridReg, denoted by $g^\theta(\cdot)$, with network parameter $\theta=\{\theta_0, \theta_1, \theta_2\}$, \[
g^\theta(\mathbf{F}_j,\mathbf{M}_j)
=
g^{\theta_2}_{\mathrm{prd}}\!\Big(
\big\{\,g^{\theta^{(l)}_1}_{\mathrm{proj}}(\mathbf{E}^{(l)}_j)\,\big\}_{l=1}^{L}
\Big),
\qquad
\{\mathbf{E}^{(l)}_j\}_{l=1}^{L}
=
g^{\theta_0}_{\mathrm{enc}}(\mathbf{F}_j,\mathbf{M}_j),
\]
GridReg can be viewed as three sub-modules: (1) $g^{\theta_0}_{enc}(\cdot)$ extract feature \(\mathbf{E}^{(l)}_j\in\mathbb{R}^{C_l\times D_l\times H_l\times W_l}\), (2) multiple projection layers $g^{\theta_1^{(l)}}_{proj}(\cdot)$, which project 3D features at $l$-th layer of the encoder into compact 1D vectors, and $g^{\theta_2}_{dec}(\cdot)$ is used to aggregate features from encoder and projector and produce the gridded displacement \(\boldsymbol{\mu}_j \in \mathbb{R}^{3\times g_w\times g_h\times g_d}\), as shown in Figure~\ref{fig:intro-net}. }

In detail, each layer in the encoder consists of a 3D convolutional layer with ReLU activation and downsampling by a stride of 2. We set a base $C$ channels and doubled the number of channels for each subsequent layer in the encoder. 

{
Then, the features at ${l-th}$ layer from the encoder are passed through a projection layer, which includes 3D convolutional layers with fixed 8C output channels and flattening operation, resulting in a vector at $l$-th layer: \(\mathbf{P}_j^{(l)}=g^{\theta_1^{(l)}}_{proj}(\mathbf{E_j}^{(l)}) \in \mathbb{R}^{N_p\times C_p}\) with a length of $N_p = p_w\cdot p_h\cdot p_d$. 
Instead of using a symmetric decoder to reconstruct the output, as in previous work~\citep{balakrishnan2019voxelmorph}, we use a 1D cross-attention predictor $g_{dec}^{\theta_2}(\cdot)$ to progressively refine the decoder features, from bottom to top in Figure~\ref{fig:intro-net}. 
The decoder at $l$-th layer 
is expressed as:
\(
    \mathbf{{Y}}^{-(l-1)}_j = g_{dec}^{\theta_2^{(l)}}(\mathbf{\hat{Y}}^{-(l)}_j)
\)
, where $\mathbf{\hat{Y}}^{-(l)}_j={[(\mathbf{P}_j}^{(l)}), (\mathbf{Y}_j^{-(l)})] \in \mathbb{R}^{N_p\times(C_p+C_d^{-(l)})}$,} where $C_d^{-(l)}$ is decoder channel number at $l$-th layer.

\textcolor{revised}{
Let $G=g_w g_h g_d$ be the number of control points (queries) and control points coordinates stacked as $\mathbf{R}\in\mathbb{R}^{G\times 3}$ and positional encoding $\psi(\cdot)$ produce $\psi(\mathbf{R})\in\mathbb{R}^{G\times C_{pe}}$, where $C_{pe}$ is the position embedding channels. 
Given a multi-attention head and dimension in each head as $H$ and $d$, define learned projections:
$
W_Q\in\mathbb{R}^{C_{pe}\times Hd},\quad
W_K,W_V\in\mathbb{R}^{(C_p+C_d^{-(l)}))\times Hd},\quad
W_O\in\mathbb{R}^{Hd\times C_{out}^{-(l)}}.
$
Queries, keys, and values are formed by matrix multiplication:
\[
\mathbf{Q}=\psi(\mathbf{R})W_Q\in\mathbb{R}^{G\times Hd},\qquad
\mathbf{K}=\hat{\mathbf{Y}}W_K\in\mathbb{R}^{N\times Hd},\qquad
\mathbf{V}=\hat{\mathbf{Y}}W_V\in\mathbb{R}^{N\times Hd}.
\]
Reshape into $H$ heads:
\[
\mathbf{Q}\rightarrow \mathbf{Q}^{(h)}\in\mathbb{R}^{G\times d},\quad
\mathbf{K}\rightarrow \mathbf{K}^{(h)}\in\mathbb{R}^{N_p\times d},\quad
\mathbf{V}\rightarrow \mathbf{V}^{(h)}\in\mathbb{R}^{N_p\times d},\qquad h=1,\dots,H.
\]
The prodoct-attention is computed as:
\begin{equation}
\mathbf{A}^{(h)}=\mathrm{softmax}\!\left(\frac{\mathbf{Q}^{(h)}(\mathbf{K}^{(h)})^\top}{\sqrt{d}}\right)\in\mathbb{R}^{G\times N_p},
\qquad
\mathbf{Y}^{(h)}=\mathbf{A}^{(h)}\mathbf{V}^{(h)}\in\mathbb{R}^{G\times d},
\end{equation}\label{eq:att}
where $\mathrm{softmax}(\cdot)$ is applied row-wise over the $N_p$ tokens.
Concatenating heads and projecting gives the decoder output:
\(
\mathbf{Y}=\mathrm{Concat}\big(\mathbf{Y}^{(1)},\dots,\mathbf{Y}^{(H)}\big)\,W_O\in\mathbb{R}^{G\times C_{out}^{-(l)}},
\)
which is reshaped to $C_{out}^{-(l)}\times g_w\times g_h\times g_d$ at last layer of decoder.
}
\textcolor{revised}{
\paragraph{Registration models with adaptive grid sizes}
To avoid training one network per control–point grid, we implement a single model that adapts to multiple grid sizes during training and selects an optimised grid size on the validation set. We modified how the query positional encodings (PEs) are generated as described above to adapt the attention-based decoder to dynamic grid sizes. In addition, the dynamic grid implementation calculates and caches the PEs for use at runtime, enabling flexibility in target grid size.  In summary, the encoder is shared across all settings. The PE width \(C_{\!pe}\) sets the embedding feature length of each query.
 In decoder, the grid size \((g_w,g_h,g_d)\) changes only the number of queries $G=g_w g_h g_d$.
According to ~\eqref{eq:att}, the projection weights \({W}_Q,{W}_K,{W}_V,{W}_O\) and head sizes \((H,d)\) are
independent of \((g_w,g_h,g_d)\). Thus, the model can adapt to different grid sizes without additional computational cost.
}

\textcolor{revised}{
 To support a dynamic grid, we treat the grid resolution as a discrete hyperparameter and, during training, sample
\(
(g_w,g_h,g_d)\sim \mathcal{U}\big(\{(5,5,5),(8,8,8),(10,10,10),(15,15,15)\}\big),
\). The sampled grid is embedded by a position embedding $\phi(g_w, g_h, g_d)$. The cross attention module takes a position embedding as the query and produces deformation parameters $\boldsymbol{\mu} $ and $ \boldsymbol{\sigma}$ on a flexible, dynamic grid. This approach offers {computational efficiency} comparable to fixed-grid models while significantly reducing storage and deployment overhead. During inference, we select the optimal, fixed grid size based on validation metrics for each Dataset.
}

\subsection{Transformation sampling}
\label{sec.method-B}
\textbf{Bayesian grid transformer}
\textcolor{revised}{
We adopted a previously proposed probabilistic registration algorithm~\cite{gong2022uncertainty,dalca2019unsupervised} and learn a grid-based posterior over deformations, to estimate uncertainty in medical image registration. This allows the method to assess uncertainty for deformation as well as improve the performance in~\cite{gong2022uncertainty,dalca2019unsupervised}.}
\textcolor{revised}{
Thus, we add a Bayesian head after the decoder, which predicts
a mean and variance per control point,
\[
\boldsymbol{\mu}_j,\ \boldsymbol{\eta}_j \in \mathbb{R}^{3\times g_w\times g_h\times g_d}, 
\qquad \boldsymbol{\sigma}_j^2 = \mathrm{softplus}(\boldsymbol{\eta}_j),
\]
for each pair $j$. During training, we use the reparameterization trick to draw $S$ Monte–Carlo samples at the grid,
\[
\mathbf{T}^{(s)}_j \;=\; \boldsymbol{\mu}_j \;+\; \boldsymbol{\sigma}_j \odot \boldsymbol{\epsilon}^{(s)}, 
\qquad \boldsymbol{\epsilon}^{(s)} \sim \mathcal{N}(\mathbf{0},\mathbf{I}),\ s=1,\dots,S,
\]
We optimise the similarity objective with an uncertainty-weighting:
\begin{equation}
\label{eq:uncert}
\mathcal{L}_{\text{uncert}}
=\frac{1}{S}\sum_{s=1}^{S}\sum_{\mathbf{x}\in\Omega}
\frac{
\ell_{sim}\!\big((\mathbf{m_j}\circ \mathbf{T}^{(s)}_j)(\mathbf{x}),\,f_j(\mathbf{x})\big)
}{2\,\sigma_j^2(\mathbf{x})}
\;+\;
\lambda_{0}\sum_{\mathbf{x}\in\Omega}\log \sigma_j^2(\mathbf{x}),
\end{equation}
${\sigma}^2_j(\boldsymbol{x})$ is the spatially varying uncertainty, 
and $\lambda_{0}$ controls the uncertainty penalty. Intuitively, $\boldsymbol{{\sigma}}_j$ grows in ambiguous (e.g. homogeneous or artefact–corrupted) regions and reduces where correspondences are clear.  
This pathwise sampling provides an unbiased Monte Carlo estimate of the expectation, discouraging sharp, unstable deformations in homogeneous areas.
}

\textcolor{revised}{
In inference, we use only the mean field to warp the moving image,
$\mathbf{T}_{\mu,j} \equiv (\boldsymbol{\mu}_j),$
while $\boldsymbol{\sigma}_j$ is used solely for uncertainty visualisation.
}

\noindent\textbf{Transformation interpolation}
\label{sec:upsample}
 Given a predicted GDF $T(\mathbf{x}) \in \mathbb{R}^{3\times g_w\times g_h\times g_d}$, three interpolation methods are investigated to upsample the sparse GDF to image size, obtaining $T^{\uparrow}(\mathbf{x})$, where $^{\uparrow}$ represents upsampling operation. To interpolate at each voxel $\mathbf{x}_n=(x_i,y_{\nu},z_k)$, the locations of GDF are represented as $\mathbf{x}_{ijk}$.
We describe three upsampling strategies that can consistently be considered by using a basis function to convolve with the GDF, denoted as 
\begin{equation}
T^{\uparrow}(\mathbf{x}_n)=\sum_{i=1}^l\sum_{\nu=1}^l\sum_{k=1}^l T(\mathbf{x}_{i\nu k}) B_x(x_i)B_y(y_{\nu})B_z(z_k) 
\end{equation}, 
where the terms $B_x(x_i)$, $B_y(x_i)$, and $B_z(x_i)$ represent the basis functions in three dimensions. These basis functions share the same functional form across all three dimensions ($x$, $y$, and $z$).
For clarity, let us consider $B_x(x_i)$ as an example to illustrate the three interpolation methods:

{(1) Trilinear interpolation:}
The basis function in trilinear interpolation is defined as 
\(B_x(x_i)= 1-\left|x-x_i\right|\) along 1st dimension (similarly for the other two dimensions). 

{(2) Cubic B-spline interpolation:}\label{sec:bspline}
The base function of cubic B-spline interpolation is defined recursively from 0-degree to 3-degree, i.e., the 1st dimension basis function at 0-degree is defined as: 
\[
B_i^0(x) = 
\begin{cases}
1 & \text{if } t_i \leq x < t_{i+1} \\
0 & \text{otherwise}
\end{cases}, \text{where } t_0 < \cdots < t_i < \cdots < t_I
\]
 is a linearly sampled knots vector \(\mathbf{t}=[0,0,0,t_0,...,t_i,...,t_I,1,1,1]\). Then, basis functions at $p=\{1,2,3\}$ degree are defined recursively by:
\[
B_x^p(x_i)=\frac{x-t_i}{t_{i+p}-t_i}B_x^{p-1}(x_i) +\frac{t_{i+p+1}-x}{t_{i+p+1}-t_{i+1}}B_x^{p-1}(x_{i+1})
\] (similarly for the other two dimensions).

{(3) Transposed convolution upsampling:}
The 3D transposed convolutions were used to implement a Gaussian spline-based transformation resampling.
In this case, the basis function of 1st dimension is defined as: \[
B_x(x_i) = \frac{1}{\sqrt{2\pi\sigma^2}} e^{-\frac{(x-x_i)^2}{2\sigma^2}}
\],
where $\sigma$ is the standard deviation controlling the width of the kernel, which can also be empirically configured to efficiently approximate the B-spline interpolation, should it be beneficial. 
All implementations for the trilinear, B-spline interpolation and its approximation are differentiable during backpropagation, which allows the network to learn the optimal interpolation method for the task at hand~\citep{fey2018splinecnn}.

\subsection{Loss function}\label{sec.method-C}
We define image-level and grid-level similarity as the displacement field sampled at full resolution and a sparse, lower resolution using mean squared error, denoted as 
$L_{sim}^{image}$, where\begin{equation}
\mathcal{L}_{\mathrm{sim}}^{\mathrm{img}}
= \frac{1}{|\Omega|}\sum_{\boldsymbol{x}\in\Omega}
\mathrm{sim}\!\big( (\mathbf{m}_j\circ\mathbf{T}_j^{\uparrow})(\boldsymbol{x}),\; \mathbf{f}_j(\boldsymbol{x}) \big),
\end{equation}
 and $^\uparrow$ denotes an upsampling operation (described in Sec.~\ref{sec:upsample}). $Sim$ is a similarity function, such as mutual information, MSE and cross-correlation functions. Here, we adopted MSE for faster training. Additionally, $L_{sim}^{img}$ is adapted into an uncertainty-weighted loss, enabling the model to identify and adjust for regions with unreliable registration in~\eqref{eq:uncert}.

We additionally calculated Dice loss function between fixed mask $\mathbf{s}_{\mathrm{fix}}$ and warped mask from moving mask $\mathbf{s}_{\mathrm{mov}}$: 
\begin{equation}
\mathcal{L}_{\mathrm{Dice}}
= 1 - \frac{2 \sum_{\boldsymbol{x}\in\Omega}
\, \mathbf{s}_{\mathrm{fix}}(\boldsymbol{x})\;
\big(\mathbf{s}_{\mathrm{mov}}\circ\mathbf{T}^{\uparrow}\big)(\boldsymbol{x})
}{
\sum_{\boldsymbol{x}\in\Omega}\mathbf{s}_{\mathrm{fix}}(\boldsymbol{x})
\;+\;
\sum_{\boldsymbol{x}\in\Omega}\big(\mathbf{s}_{\mathrm{mov}}\circ\mathbf{T}^{\uparrow}\big)(\boldsymbol{x})
+\epsilon }.
\end{equation}
    to measure the overlap between the predicted segmentation mask and the ground truth segmentation mask.  
A bending energy loss is defined as:
\begin{equation}
\mathcal{L}_{bend} = \frac{1}{X} \sum_{i} \left( \left| \nabla^2 \mathbf{t}_i \right|^2 + 2 \sum_{\rho \neq \varrho} \left| \frac{\partial^2 \mathbf{t}_i}{\partial \rho \partial \varrho} \right|^2 \right)
\end{equation},
where $\nabla^2$ is the Laplacian operator, and $\mathbf{t}_i$ is the displacement field at voxel location $i$,  \( \rho, \varrho \in \{x, y, z\} \) denote 3 spatial dimensions. The bending energy loss penalises the second derivative of the displacement field, encouraging the network to generate smoother deformation fields~\cite{rueckert1999nonrigid}.

The final loss can be written as: 
\(    \mathcal{L} = \lambda_1 \mathcal{L}_{uncertainty} + \lambda_2 \mathcal{L}_{DSC} + \lambda_3 \mathcal{L}_{bend}\)
,
where $ \lambda_1, \lambda_2$ and $\lambda_3$ are hyperparameters that control the relative importance of the different loss terms. 
\textcolor{revised}{Dice loss is only included when segmentation masks for training data are available, i.e., Dataset~1 and Dataset~2.}

\section{Experiments}
\subsection{Datasets}
\begin{figure}
    \centering
    \includegraphics[width=0.8\linewidth]{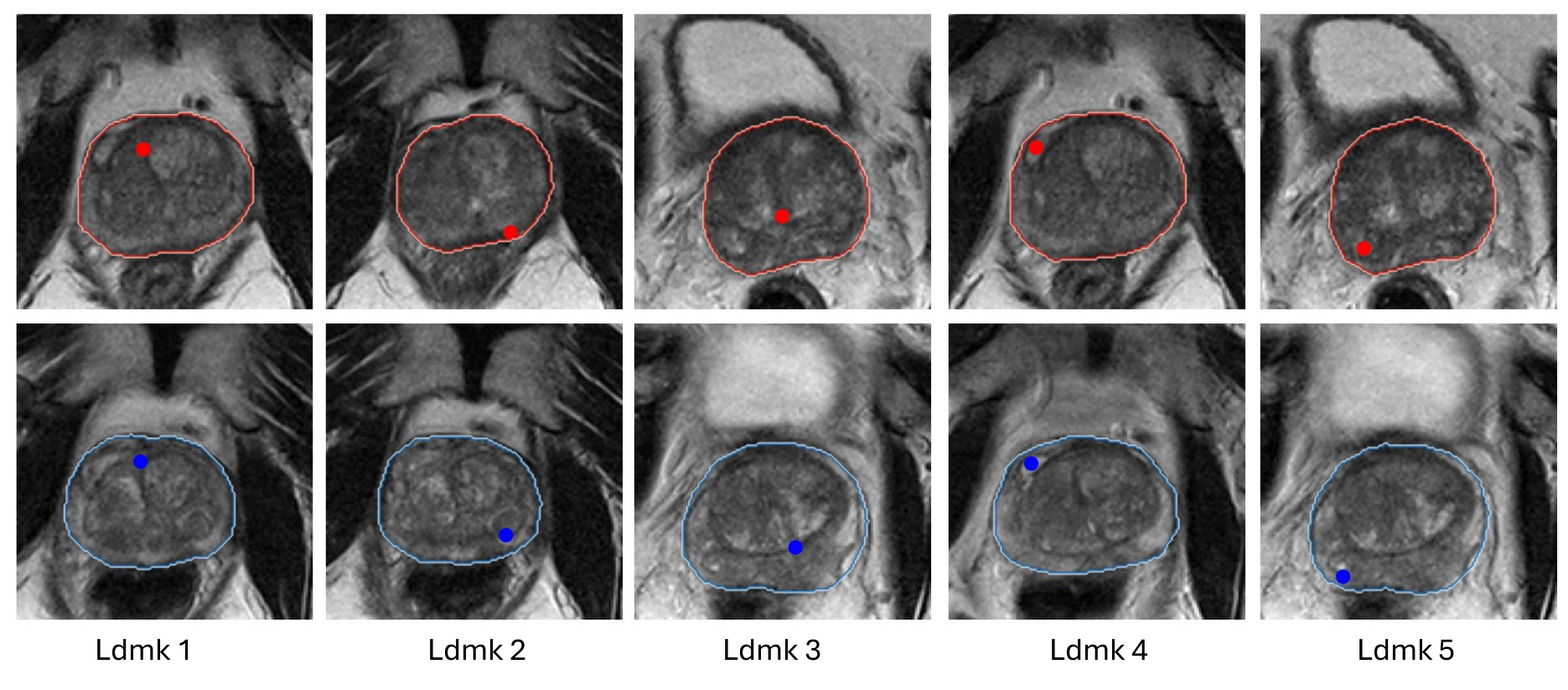}
    \caption{Examples of anatomical landmarks annotated for the same patient at two different time points. The first and second rows show 5 landmarks at the first and second time points, respectively.}
    \label{fig:prostateldmk}
\end{figure}

\textbf{Dataset 1:} \textcolor{revised}{Longitudinal prostate T2-weighted MR images were acquired from 86 patients at University College London Hospitals NHS Foundation. All patients provided written consent.
Among them, 60 patients had 2 visits, 8 patients had 3 visits, and 18 patients had 4 visits, yielding a total of 216 T2-weighted volumes.}
This dataset was used to perform \textit{intra-patient} registration. 
All the image and mask volumes were resampled to $0.7\times0.7\times0.7$ $mm^3$ isotropic voxels. The intensity of all image volumes is normalised to a range of $[0,1]$. To facilitate computation, all images and masks were also cropped from the centre of volume to $128\times128\times102$ voxels with preserved prostate glands. 
The data set was divided into 70, 6 and 10 patients for training, validation, and \textcolor{revised}{holdout test sets}, composing 156, 32 and 38 intra-patient pairs, respectively. For images in the \textcolor{revised}{holdout test set}, pairs of corresponding anatomical and pathological landmarks are manually identified on moving and fixed images \textcolor{revised} {by a radiologist with more than 5 years of experience in reading prostate cancer mpMR images.} Examples are shown in Figure~\ref{fig:prostateldmk}, including patient-specific fluid-filled cysts, calcification, and centroids of zonal boundaries. The landmarks differ from patient to patient because of the complex anatomical structure of the prostate.

\textbf{Dataset 2:} We used pelvis data from 850 prostate cancer patients at 
University College London Hospital 
for \textit{inter-patient} registration. 
Details of the data can be found in ~\citep{hamid2019smarttarget,simmons2017picture,orczyk_proraft,dickinson_index,bosaily_promis}. All patients provided written consent, and ethics approval was granted as per trial protocols in~\citep{hamid_smarttarget}. We use 313, 58, and 69 patients for training, validation and \textcolor{revised}{holdout test sets, respectively. This dataset spans several clinical trials conducted at UCLH, with images acquired from various clinical scanners. The patient group includes both biopsy and therapy cases.}
Specifics on vendors and imaging protocols for each study can be found in~\citep{yan2024combiner}.
All patients provided written consent and ethics approval was granted as per trial protocols~\citep{hamid_smarttarget}. 
\textcolor{revised}{We perform inter-patient registration on Dataset~2. The 850 patients (30 patients have multiple T2 modalities) were randomly partitioned into train, validation and holdout test sets, on patient-level, resulting in 626, 116 and 118 patients, respectively. Within each set, image pairs are randomly formed and sampled without replacement, yielding 313, 58, and 59 pairs for train, validation, and holdout test sets, respectively.  }
T2-weighted images have in-plane dimensions ranging from 180$\times$180 to 640$\times$640, with a resolution of 1.31$\times1.31 mm^2$ to 0.29 $\times 0.29 mm^2$, and slice thickness between 0.82 and 1 mm. All image modalities were resampled to isotropic voxels of 1 $\times 1 \times 1 mm^3$ using linear interpolation, with intensity normalised between 0 and 1 per modality. \textcolor{revised}{Prostate masks are labelled by radiologists with more than 5 years of experience in interpreting prostate mpMR images. Inter-subject corresponding landmarks are unavailable in this dataset.}

\begin{figure}[htb]
    \centering
    \includegraphics[width=0.7\linewidth]{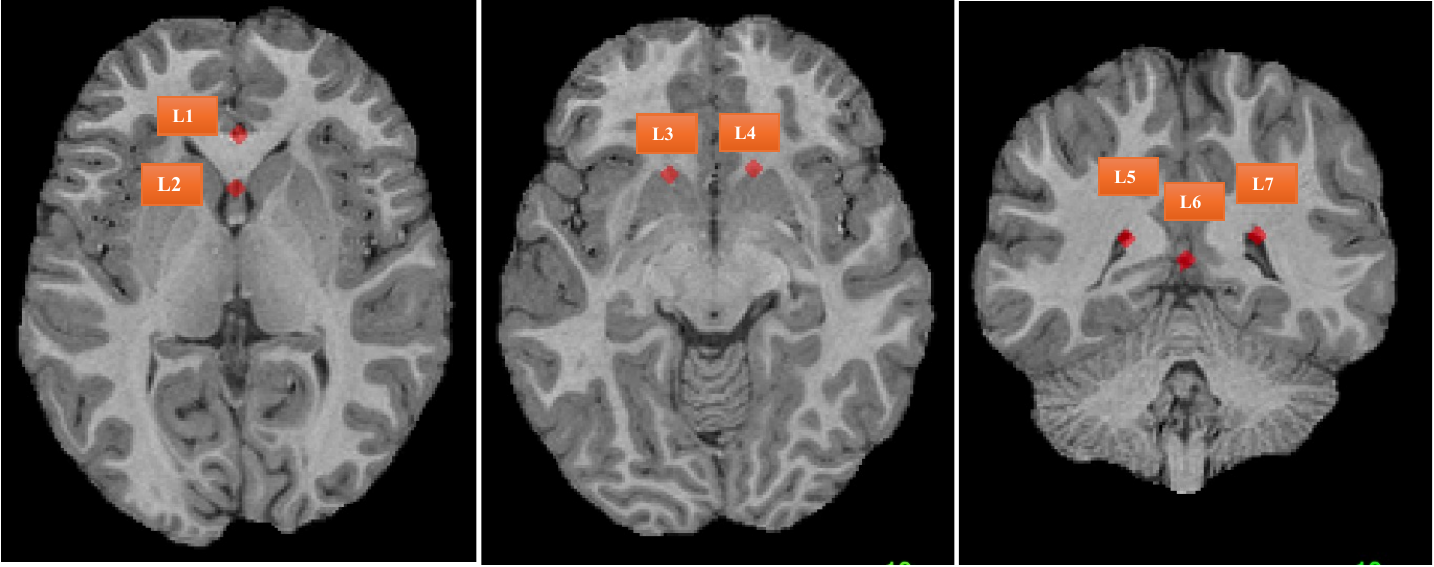}
    \caption{Examples of 7 landmarks labeled by experts in brain dataset.}
    \label{fig:brainldmk}
\end{figure}

\textbf{Dataset 3:} We used a public brain \textcolor{revised}{MR dataset LUMR from Learn2Reg~\citep{hering2022learn2reg} to perform \textit{inter-patient} registration}, which contains 3384 training and 40 validation sets, respectively. All image modalities were resampled to isotropic voxels of 1 $\times 1 \times 1 mm^3$ using linear interpolation, with intensity normalised between 0 and 1 per modality.
We split the dataset into 3000, 384 and 40 for training, validation and holdout test, respectively. We used FreeSurfer~\citep{fischl2012freesurfer} to generate brain zonal segmentation, and 7 landmarks were labeled by experienced experts for holdout test data as shown in Figure~\ref{fig:brainldmk}.
No segmentation label was provided for training; thus, the dice loss was not used during training for this dataset. 

\textcolor{revised}{It is worth noting that only training and validation sets are used in updating weights and tuning hyperparameters. holdout test sets are only used for model inference; all the results in this paper are based on the holdout test set.}

\subsection{Comparison and ablation studies}
\subsubsection{Comparison studies} 
To rigorously assess the performance of the proposed method, we conducted extensive experiments across three datasets, comparing against representative state-of-the-art approaches: dense deformation field estimation and keypoint-based methods. 
For dense-grid-based approaches, we selected VoxelMorph~\citep{balakrishnan2019voxelmorph}, a CNN-based method that predicts voxel-wise displacement fields using U-Net architectures, and TransMorph~\citep{chen2022transmorph}, a recent transformer-based variant that leverages self-attention mechanisms to model long-range spatial dependencies. 
In the keypoint-based category, we benchmarked against KeyMorph~\citep{evan2022keymorph}, which employs a sparse set of automatically detected anatomical keypoints to drive the registration.
To further elucidate the efficiency advantages of our method, we performed a focused architectural comparison with KeyMorph, analysing the relationship between model complexity (quantified by encoder depth) and registration accuracy. 
\textcolor{revised}{We included a strong iterative baseline: ANTs SyN(diffeomorphic) ~\cite{balakrishnan2019voxelmorph} with matched pre-processing and masks, tuned on the validation set. }

\subsubsection{Ablation studies}
We evaluated the significance of key components by isolating these elements. We identified their individual contributions to model performance and validated the effectiveness of our architectural decisions.
Ablation studies involved the following experiments:
(1) Projected-skip connection: We compared the performance of the proposed method with and without the skip-projection connection, which is used to project 3D features into 1D vectors.
(2) Grid size: We investigated the impact of varying grid sizes on registration performance, specifically using grid sizes of 5, 8, 10, and 15 in each dimension.
(3) Encoder channel $g_{enc}^{\theta}$: We assessed the influence of different base channels (8, 16, and 32) in the encoder on registration performance.
(4) Hyperparameters in loss function: We evaluated the effect of different hyperparameters in the loss function.
\textcolor{revised}{(5) Bayesian head: We evaluate how the Bayesian head affects registration accuracy and deformation regularity.}

\subsection{{Network and baselines implementations}}
We implemented all experiments with PyTorch 2.0 under CUDA 12.4 on NVIDIA-SMI 550.107.02 with 32GB memory. The optimiser was Adam, with a learning rate of $10^{-4}$ and a batch size of 4. The network was trained for 300 epochs. All comparison methods were trained with their official implementation.

\subsection{Evaluation metrics}
\paragraph{Registration accuracy metrics}
The Dice similarity coefficient (DSC) was used to measure the overlap between the fixed and warped prostate masks. Centroid distance is adopted to evaluate the mean Euclidean distance between corresponding centroids:
\(
D_{\mathrm{CD}}=\frac{1}{K}\sum_{i=1}^{K}\left\lVert \mathbf{c}^{\,\mathrm{fix}}_i-\mathbf{c}^{\,\mathrm{warp}}_i \right\rVert_2,
\)
where \(\mathbf{c}^{(\cdot)}_i\) denotes the \(i\)-th landmark centre or region of interest mask centre in the fixed/warped image, denoted as $D^{\mathrm{ldmk}}_{\mathrm{CD}}$ and $D^{\mathrm{msk}}_{\mathrm{CD}}$, respectively.  the centroids of the fixed and warped ROI masks, computed as
\(
\mathbf{c}^{(\cdot)}=\frac{\sum_{\mathbf{x}\in\Omega}\mathbf{x}\, s^{(\cdot)}(\mathbf{x})}{\sum_{\mathbf{x}\in\Omega} s^{(\cdot)}(\mathbf{x})},
\)
with \(s(\mathbf{x})\in\{0,1\}\) the mask value at voxel \(\mathbf{x}\).

\textcolor{revised}{
\paragraph{Jacobian-based deformation regularity.}
Let $\phi:\Omega\subset\mathbb{R}^3\rightarrow\mathbb{R}^3$ denote the estimated spatial transformation. The Jacobian matrix of $\phi$ at voxel $x$ is
$
J_\phi(x)=\nabla \phi(x)\in\mathbb{R}^{3\times 3}.
$
We quantify local volume change using the log-Jacobian determinant
$
\log\det(J_\phi(x)),
$
where values near $0$ indicate near-incompressible mappings, positive values indicate local expansion, and negative values indicate local contraction, denoted as $\log detJ$. We also reported the folding rate, defined as the percentage of voxels with a negative Jacobian determinant,
$
\det(J_\phi(x))<0,
$
which corresponds to non-invertible (folded) transformations.
}

\textcolor{revised}{
For the overall methods comparison, we compared GridReg to each baseline on the same cases using paired, one-sided tests and controlled multiplicity within each dataset $\times$ metric family using Benjamini–Hochberg FDR correction~\cite{benjamini1995controlling} at 5\%; we report $q$-values. For the ablation study, as they are independent one-by-one comparisons, we performed a paired t-test with a significance level $\alpha=0.05$.}

\section{Results}
\subsection{Comparison results}
\textcolor{revised}{In Table~\ref{tab:results}, ANTs SyN was $~40\times$ slower than learning-based methods and yielded lower Dice/centroid distance. }
VoxelMorph and Transmorph, which used DDF, emphasised global shape registration and achieved much larger landmark TRE (8.76mm and 7.73mm) in Table~\ref{tab:results} with Dataset 1. 

\begin{figure*}[htb]
    \centering
    \includegraphics[width=0.95\linewidth]{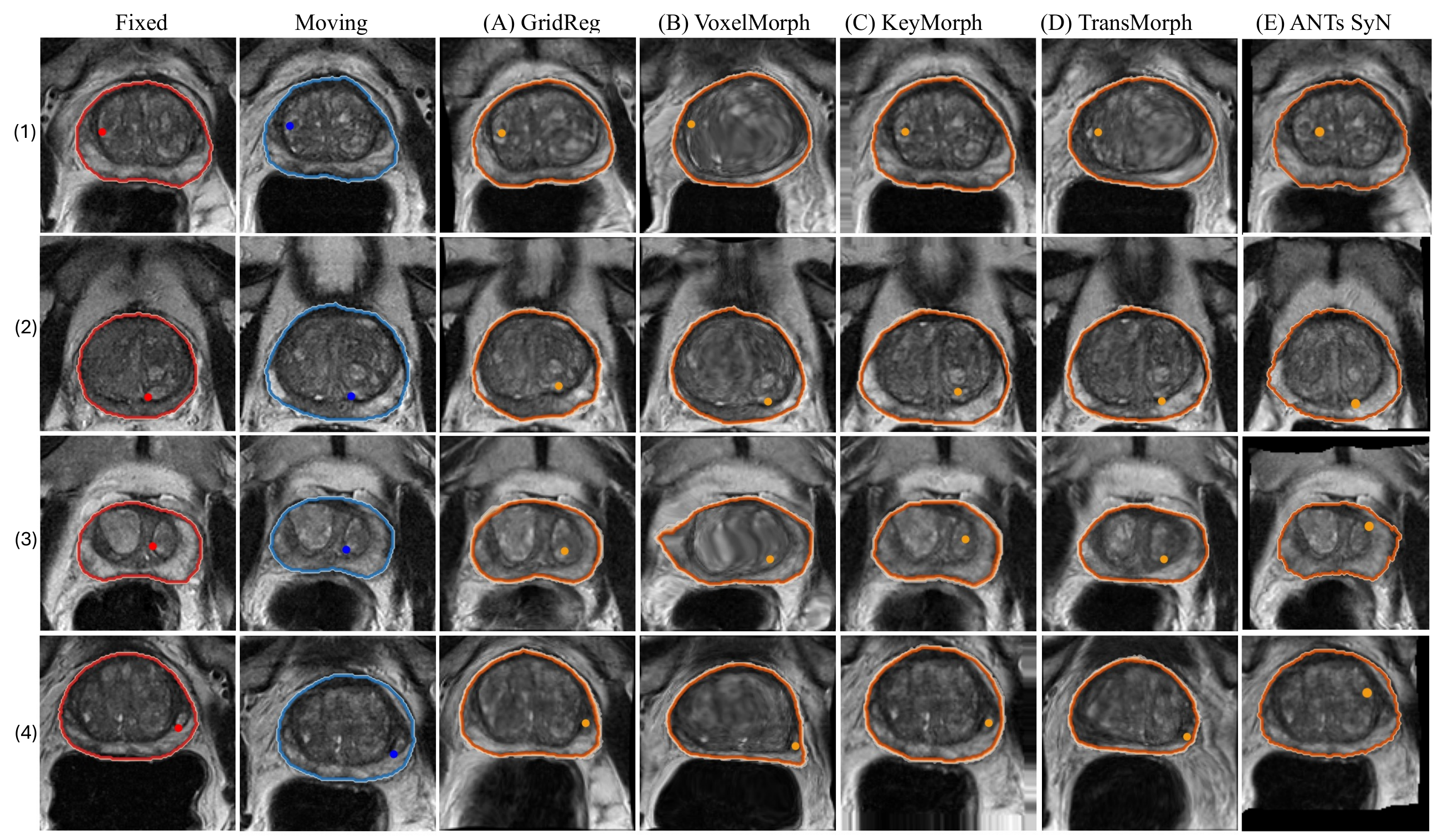}
    \caption{Visualisation of prostate registration. The first column and the second column show four examples of the fixed and moving images, respectively. The next four columns show the registration results from\textcolor{revised}{ (A)GridReg, (B) VoxelMorph, (C) KeyMorph, (D)TransMorph and (E) ANTs SyN, respectively. }}
    \label{fig:prostate_visual}
\end{figure*}

\begin{figure*}[!htb]
    \centering
    \includegraphics[width=0.9\linewidth]{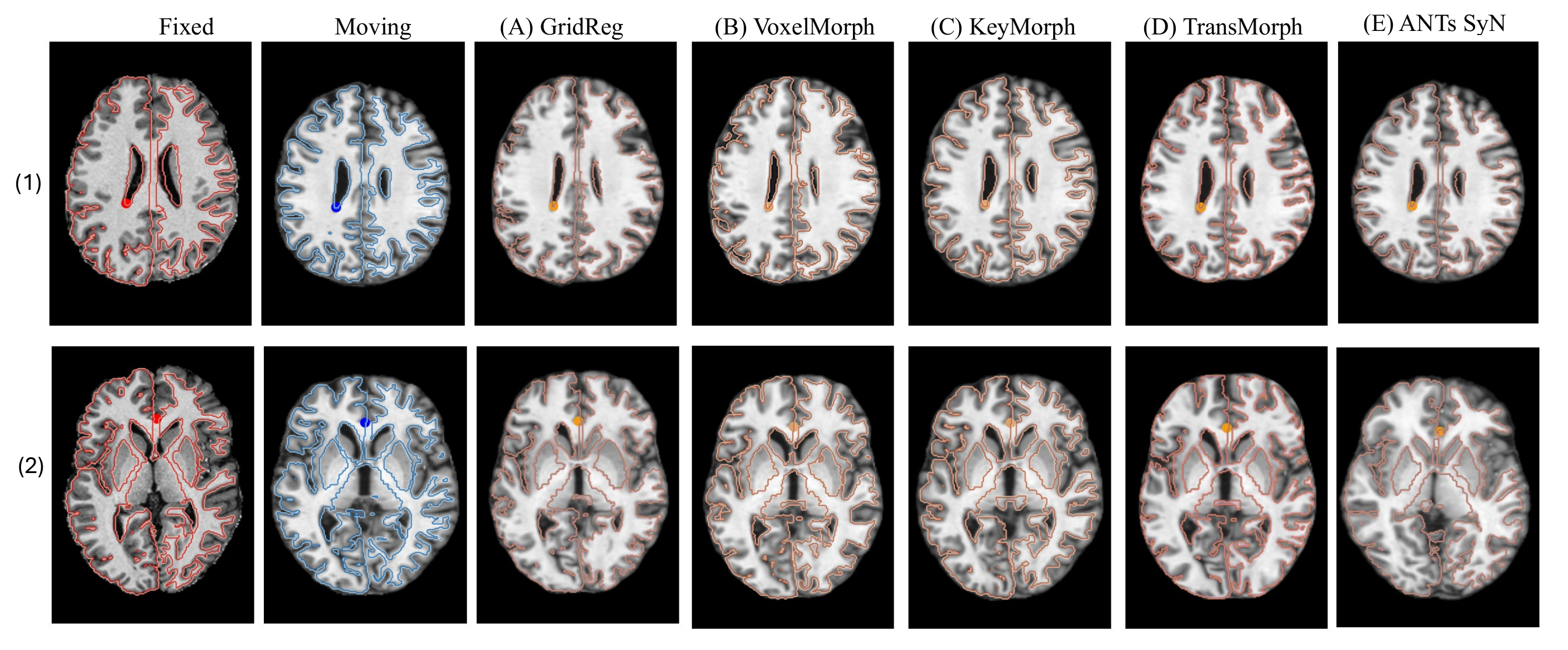}
    \caption{Visualisation of brain registration. The first column and the second column show four examples of the fixed and moving images, respectively. The next four columns show the registration results from \textcolor{revised}{(A)GridReg, (B) VoxelMorph, (C) KeyMorph, (D)TransMorph and (E) ANTs SyN, respectively.}}
    \label{fig:brain_visual}
\end{figure*}

For Dataset 2,  GridReg achieved a DSC of 0.78, slightly lower than KeyMorph's score of 0.79 but higher than TransMorph and VoxelMorph. Both GridReg and KeyMorph achieved comparable performance in mask-based CD for the pelvis, with GridReg at 4.52 mm and KeyMorph at 4.14 mm, suggesting reliable mask alignment. 
For Dataset3, GridReg maintained a DSC of 0.75, outperforming VoxelMorph and TransMorph. GridReg matched KeyMorph on mask-based CD (2.97 mm vs 3.05 mm, $p=0.832$) and trailed slightly on Dice (0.77 vs 0.78, $p=0.471$) but using less computation cost. These results suggested that when data features were rich in correspondence, architectural or transformation constraints mattered less, and performance differences were narrow. 
\textcolor{revised}{
Figure~\ref{fig:prostate_visual} showed prostate registration, where large homogeneous regions made dense DDF models (e.g., VoxelMorph, TransMorph) more prone to physically unlikely local distortions, despite comparable Dice scores. Despite fine-tuning the hyperparameters for bending energy loss, the network struggled to preserve the prostate structure. This highlights the importance of method selection in registering various medical images. 
In contrast, Figure~\ref{fig:brain_visual} shows two fixed–moving brain pairs (first two columns) and the corresponding results from four methods, i.e. GridReg, VoxelMorph, KeyMorph and TransMorph, respectively, without implausible deformation readily visible across all methods. 
}

\begin{table*}[htb]
\centering
\caption{\textcolor{revised}{Performance and memory across methods. Values are mean $\pm$ SD over identical holdout test cases. Proposed methods appear in the last columns (GridReg). For each dataset, we compared GridReg to each baseline using paired, one-sided t-tests and controlled multiplicity within the family using BH-FDR ($5\%$); we report q-values. Bold indicates the best mean. $\ddag$ denotes GridReg is significantly better than all baselines ($q<0.05$); $\dag$ denotes it is significantly better than three of baselines ($q<0.05$).} Slash $/$ indicates that a result is not available.}
    \resizebox{\linewidth}{!}{
\begin{tabular}{cc|c>{\color{revised}}ccccc}
    \hline
    \textbf{Dataset} & \textbf{Metric} & \textbf{Original} & \textbf{{ANTs SyN}} & \textbf{VoxelMorph} & \textbf{TransMorph} & \textbf{KeyMorph} & \textbf{GridReg(Ours)} \\
    \hline
    \multirow{4}{*}{\textbf{Prostate}}
    &Dice$\uparrow$ & 0.70$\pm$0.10 &$0.81\pm0.05$& 0.86$\pm$0.02 & 0.84$\pm$0.03 & 0.86$\pm$0.05 & \textbf{0.88}$\pm$0.02{$^\ddag$} \\

    &$D^{ldmk}_{CD}\downarrow$(mm) & 9.60$\pm$4.09 & 9.09$\pm$8.09 & 8.76$\pm$8.59 & 7.73$\pm$6.98 & \textbf{3.98$\pm$2.14} & 4.54$\pm$2.77{$^\dag$} \\
  
    &$D^{msk}_{CD}\downarrow$(mm) & 8.76$\pm$4.04 & 3.38$\pm$0.78 & 2.70$\pm$0.54 & 3.05$\pm$0.84 & 2.56$\pm$0.66 & \textbf{1.87$\pm$1.16}{$^\ddag$} \\

    &GPU(Mb) & / & /& 208 & 377 & 274 & \textbf{146} \\
    \hline
    \multirow{3}{*}{\textbf{Pelvis}}
    &Dice$\uparrow$ & $0.53\pm0.17$ & 0.69$\pm0.09$ & 0.73$\pm$0.16 & 0.73$\pm$0.09 & \textbf{0.79$\pm$0.07} & 0.78$\pm$0.09{$^\dag$} \\
    
    &$D^{msk}_{CD}\downarrow$(mm) & 10.65$\pm5.85$ & 6.03$\pm2.78$ & 5.99$\pm5.68$ & 4.47$\pm2.95$ & \textbf{4.14$\pm$2.47} & 4.52$\pm$2.88 {$^\dag$} \\
    &GPU(Mb) & / & / & 168 & 333 & 225 & \textbf{128} \\
    \hline
    \multirow{4}{*}{\textbf{Brain}}
    &Dice $\uparrow$& 0.70$\pm$0.05 &  0.69$\pm0.12$& \textbf{0.78$\pm$0.07} & 0.77$\pm$0.05 & 0.77$\pm$0.06 & \textbf{0.78$\pm$0.06} \\
    &$D^{ldmk}_{CD}\downarrow$(mm) & 8.67$\pm$6.58 &$8.92\pm6.10$& 8.45$\pm$6.64 & 8.70$\pm$6.64 & 8.35$\pm$6.85 & \textbf{8.21$\pm$6.58}{$^\dag$} \\
    &$D^{msk}_{CD}\downarrow$(mm) & 4.16$\pm$1.94 & 3.98$\pm1.79$ & 3.11$\pm$1.73 & 3.80$\pm1.14$ & \textbf{2.97$\pm$1.72} & 3.05$\pm$1.47 \\
    &GPU(Mb) & / & /& 317 & 486 & 378 & \textbf{240} \\
    \hline
    
    \multicolumn{2}{c|}{Trainable Parameters (bytes)} & / & / & 524181 & 41476659 & 1410329 &252960 \\
   
    \multicolumn{2}{c|}{Inference Time (case/s)} & / &8.08 & 0.23 & 0.25 & 0.24 & 0.23 \\
    
    \hline
    \end{tabular}
    }
\label{tab:results}
\end{table*}

\begin{table*}[b]
    \centering
     \caption{KeyMorph vs. GridReg: Performance comparison across varying Encoder layer counts (Dataset 1). Results show Dice, $D^{ldmk}_{CD}$(mm), and inference memory cost. \textcolor{revised}{An asterisk ($*$) indicates a statistically significant improvement over the comparator method at the same layer count ($p<0.05$).}} 
    \adjustbox{max width=\linewidth}{
        \begin{tabular}{c|ccc|ccc}
            \hline
            \multirow{2}{*}{Layers} & \multicolumn{3}{c}{KeyMorph} & \multicolumn{3}{|c}{GridReg} \\
            \cline{2-4} \cline{5-7}
            & Dice & $D^{ldmk}_{CD}$(mm) & Cost(Mb) & Dice & $D^{ldmk}_{CD}$(mm) & Cost(Mb) \\ \hline

            4 layers & 0.54$\pm$0.14 & 14.74$\pm$9.46 & 169 & 0.88$\pm$0.03\textcolor{revised}{$^{*}$} & 6.05$\pm$3.43\textcolor{revised}{$^{*}$} & 145 \\

            5 layers & 0.56$\pm$0.14 & 14.69$\pm$12.40 & 171 & 0.88$\pm0.02$\textcolor{revised}{$^{*}$} & 4.54$\pm$2.77\textcolor{revised}{$^{*}$} & 146 \\

            6 layers & 0.58$\pm$0.19 & 10.97$\pm$7.52 & 174 & 0.89$\pm$0.03\textcolor{revised}{$^{*}$} & 3.97$\pm$2.38\textcolor{revised}{$^{*}$} & 147 \\
  
            7 layers & 0.82$\pm$0.18 & 4.42$\pm$3.24 & 182 & 0.89$\pm$0.02\textcolor{revised}{$^*$} & 4.59$\pm$3.09 & 148 \\
        
            8 layers & 0.84$\pm$0.07 & 4.33$\pm$2.84 & 195 & 0.89$\pm$0.04\textcolor{revised}{$^{*}$} & 4.25$\pm$2.46 & 152 \\
     
            9 layers & 0.86$\pm$0.05 & 3.98$\pm$2.14\textcolor{revised}{$^*$} &208 & 0.90$\pm$0.03\textcolor{revised}{$^{*}$} & 4.64$\pm$2.82 & 157 \\
            \hline
        \end{tabular}
        }
       
        \label{tab:comparision_key_grid}
\end{table*}

As shown in the last two rows of Table~\ref{tab:results}, 
\textcolor{revised}{The inference time were: GridReg 0.23 s, VoxelMorph 0.23 s, KeyMorph 0.24 s, TransMorph 0.25 s, respectively. Although GridReg used ~2×–163× fewer parameters than the baselines, single-image inference time is similar because GPU execution is dominated by reading the input images and copying them from memory to the compute units. Our method substantially reduces the number of parameters, enabling deployment on smaller GPUs, but it does not noticeably reduce inference time due to the nature of GPU computing.} Keymorph achieved slightly lower DSC than the proposed method. It achieved higher accuracy in terms of centroid distances of landmarks, but it required more computation cost, as shown in Table~\ref{tab:comparision_key_grid}.
We compared KeyMorph and GridReg with varying encoder layers. KeyMorph’s DSC improved from 0.54 at 4 layers to 0.86 at 9 layers, requiring deeper networks for improved performance. GridReg consistently outperforms KeyMorph across all configurations, with DSC from 0.88 to 0.90. This highlighted GridReg’s ability to maintain high performance with fewer parameters. Though GridReg achieved slightly lower landmark distances compared to KeyMorph with more than 7 layers, it achieved more stable CD across all layers, further demonstrating its efficiency and reduced computation cost.
KeyMorph demonstrated good performance on Dataset 3 (brain imaging), where structural consistency supports its keypoint-based design, but performed less effectively on prostate datasets, where such consistency is lacking, as shown in Table~\ref{tab:results}.
\\

\subsection{Ablation studies}\label{sec:ablation}

\begin{table}[htb] 
 \caption{Ablation study of proposed method with varying configuration of grid number, base channel of encoder, upsampling method and usage of projection module. \textcolor{revised}{The ablation studies are performed on holdout test set in Dataset~1. The symbol * represents that the best results have a statistical difference between \textbf{Config 4} and other methods (all $p-values<$0.050).} }
    \centering
    \resizebox{\textwidth}{!}{
    \begin{tabular}{c|ccccc|c|c|c}
        \hline
        Setting & Grid & Base-NC & Upsample & Projector & Bayesian & Dice & $D_{CD}^{lmdk}$(mm) & Cost(Mb) \\
        \hline
        Config 1 & 5 & 32 & trilinear & $\checkmark$ & $\checkmark$ & 0.86$\pm$0.05$\textcolor{revised}{^{*}}$ & 5.29$\pm$2.85 & 146 \\
        Config 2 & 10 & 32 & trilinear & $\checkmark$ & $\checkmark$ & 0.89$\pm$0.03 & 4.89$\pm$3.19 & 148 \\
        Config 3 & 10 & 8 & trilinear & $\checkmark$ & $\checkmark$ & 0.87$\pm$0.04 & 5.88$\pm$3.27 & 144 \\
        \textbf{Config 4} & 10 & 16 & trilinear & $\checkmark$ & $\checkmark$ & 0.88$\pm$0.02 & 4.54$\pm$2.77 & 146 \\
        Config 5 & 10 & 16 & trilinear & $\times$ & $\checkmark$ & 0.86$\pm$0.03$\textcolor{revised}{^*}$ & 6.06$\pm$3.23$^{\textcolor{revised}{*}}$ & 145 \\
        Config 6 & 10 & 16 & deconv & $\checkmark$ & $\checkmark$ & 0.89$\pm$0.04 & 4.18$\pm$2.88 & 171 \\
        Config 7 & 10 & 16 & bspl & $\checkmark$ & $\checkmark$ & 0.89$\pm$0.03 & 4.45$\pm$2.95 & 290 \\
        Config 8 & 10 & 16 & trilinear & $\checkmark$ & $\times$ & 0.88$\pm$0.03 & 4.90$\pm$3.65 & 146 \\
        Config 9 & 15 & 16 & trilinear & $\checkmark$ & $\checkmark$ & 0.90$\pm$0.03$^{\textcolor{revised}{*}}$& 5.29$\pm$3.09 & 148 \\ 
        \hline
    \end{tabular}}
   
    \label{tab:ablation}
\end{table}

We conducted ablation studies to assess the effectiveness of the proposed method on Dataset~1. (1), removing the skip-projection module resulted in a lower DSC of 0.86 compared to GridReg ($p=2.02\times 10^{-234}$). The skip-projection and decoder refine the deformation field, capturing both global and local information for better accuracy.  
(2), varying the grid control points showed that (10,10,10) control points achieved the best accuracy (Table~\ref{tab:ablation}). 
(3), increasing base channels improved accuracy, with 32 channels performing best Dice, though the difference from 16 channels was not significant (DSC $p=0.053$).  
(4), cubic B-spline (via transposed convolution) achieved higher centroid distances (p = 0.800) than trilinear interpolation, while trilinear was slightly faster. The transposed-convolutional B-spline is markedly more efficient than the zero-insertion B-spline, which inflates memory by expanding the grid with zero before filtering.
\begin{figure}[htb]
    \centering
    \includegraphics[width=\linewidth]{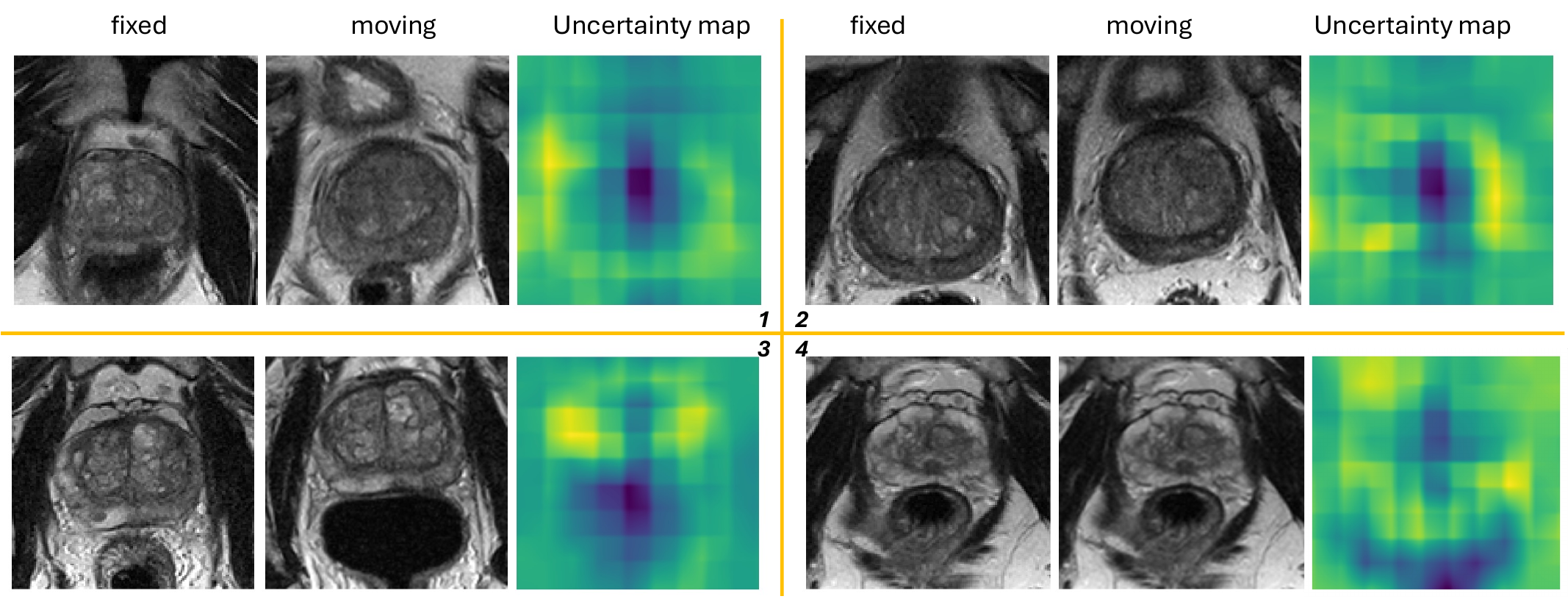}
    \caption{This figure shows four examples of uncertainty maps generated by the Bayesian Grid transformer. A lighter area in the Uncertainty map represents lower uncertainty, while a darker area represents higher uncertainty.}
    \label{fig:uncertainty}
\end{figure}

\begin{figure}[htb]
    \centering
    \includegraphics[width=\linewidth]{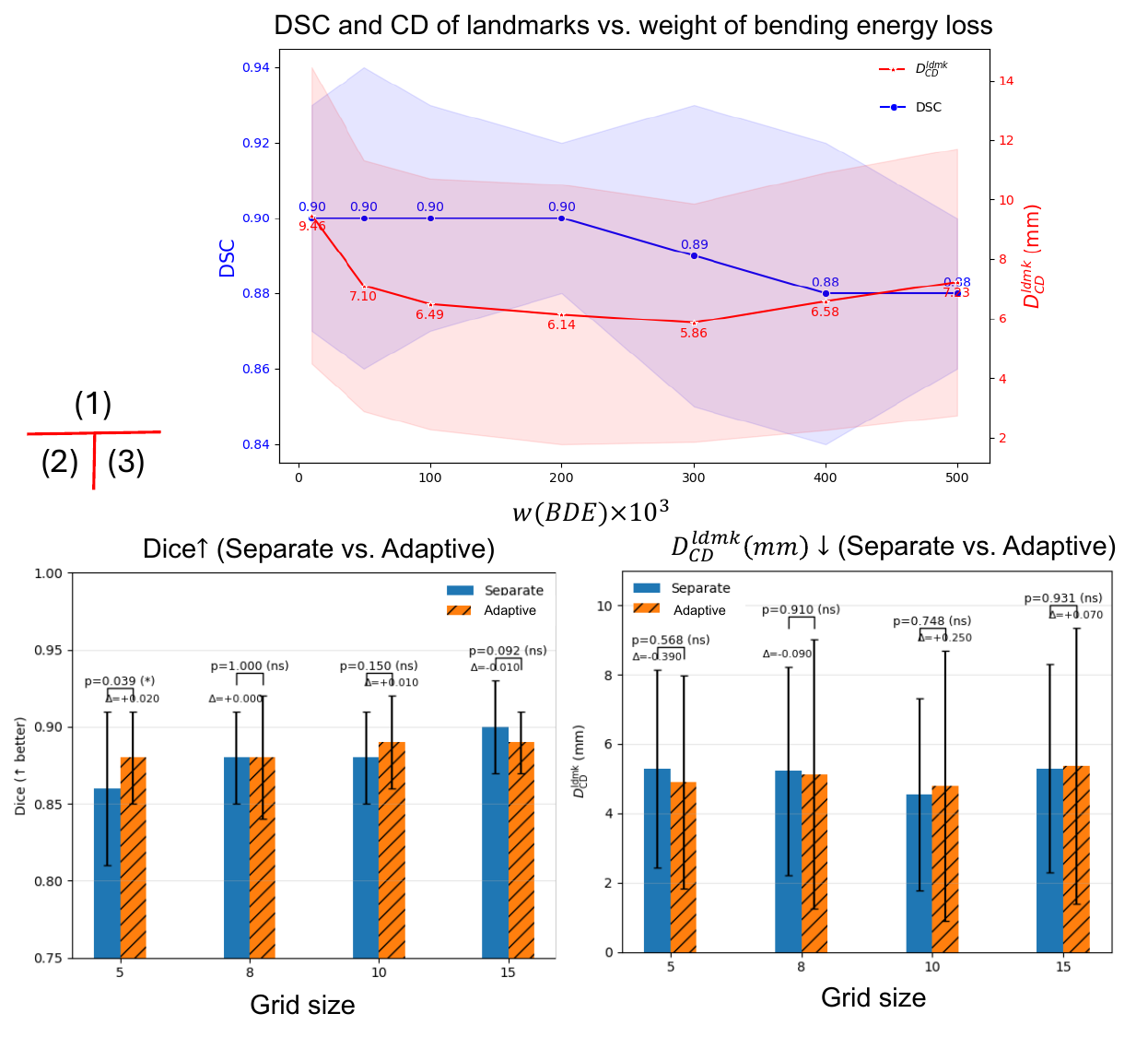}
    \caption{\textcolor{revised}{(1)Effect of Bending Energy Weight on Segmentation Accuracy (DSC) and Contour Distance (CD); (2)-(3)Comparisons on Dataset1 showing Dice and centroid distance between separately trained and auto-adapted models with the same grid size range from 5 to 15.}}
    \label{fig:auto}
\end{figure}

\textcolor{revised}{(5), The Bayesian head has been used as a regulariser in prior work~\cite{gong2022uncertainty,dalca2019unsupervised}. In our experiments, adding this head did not significantly improve registration accuracy ($p=0.815$). Likewise, Jacobian-based regularity metrics showed no significant differences, although we observed a near-significant trend on Brain (Prostate: $p=0.513$; Brain: $p=0.063$; Table~\ref{tab:jacobian_stats}).
Figure~\ref{fig:uncertainty} presents two representative cases from the holdout test set, showing the fixed image, moving image, and corresponding uncertainty maps ($\boldsymbol{\sigma}^2$), where homogeneous regions exhibit higher uncertainty than high-contrast boundaries. 
}
(6), Figure~\ref{fig:auto}~(1) shows that balancing bending energy loss weight ($2\times10^5$) optimally reduces distortion and aligns regions of interest, achieving DSC=0.90 and $D_{CD}^{ldmk}=6.14$mm.
\textcolor{revised}{
Figure~\ref{fig:auto}~(2–3) compares the separately trained and the grid-variable model across different grid sizes on Dataset 1. Using the same grid sizes, performance is similar: no significant differences were detected ($p>0.205$). The only visible gap is at grid = 5, where the grid-adaptive model shows a $0.39 mm$ lower centroid distance (CD) than that of the model with prefixed grid size ($p=0.570$) and 0.02 higher Dice than the prefixed grid size ($p=0.043$).}

\begin{table}[!ht]
\centering
\caption{\textcolor{revised}{Jacobian statistics across methods. $\log\det(J)$ is reported as mean$\pm$standard deviation; folding rate is the percentage of voxels with $\det(J)<0$. For each dataset, we report paired, one-sided t-tests with BH-FDR(5\%) with confidence $\alpha=0.05$. $\dagger$ and $^*$ denote GridReg Bayesian and GridReg is significantly better than other methods, respectively.}}
\resizebox{\linewidth}{!}{
\begin{tabular}{llccccc}
\toprule
Dataset & Metric & GridReg Bayesian & GridReg & KeyMorph & VoxelMorph & TransMorph \\
\midrule
\multirow{2}{*}{Prostate} 
& $\log\det(J)$  & $0.0352\pm0.0193^\dagger$ & $0.0446\pm0.0228^*$ & $13.3064\pm32.1452$ & $1.9258\pm1.0610$ & $0.0975\pm0.0532$ \\
& $\det(J)<0$ (\%)            & $0$               & $0$               & $3.8$               & $0.4$             & $0.03$ \\
\midrule
\multirow{2}{*}{Brain}
& $\log\det(J)$  & $0.0022\pm0.0015^\dagger$ & $0.0038\pm0.0014^*$ & $55.2012\pm133.4529$ & $1.0793\pm0.3254$ & $15.9102\pm6.5335$ \\
& $\det(J)<0$ (\%)            & $0$               & $0$               & $11.0$              & $0.2$             & $4.5$ \\
\bottomrule
\end{tabular}
}

\label{tab:jacobian_stats}
\end{table}

\section{Discussion}
\textcolor{revised}{Learning-based registration trades hand-crafted objectives for data-driven models that learn complex, non-linear correspondences and provide single-pass fast inference with sufficient data requirements. Besides, with the domain-shift strategy, it could transfer across domains~\cite{zheng2021unsupervised}. In data-scarce settings, classical methods with strong inductive priors may remain competitive. With moderately more data, as demonstrated in all our experiments, however, learning-based registration attains state-of-the-art accuracy with substantially lower runtime. }Across three datasets, our results highlight complementary strengths and failure modes of current 3D registration strategies, and motivate design choices behind GridReg. 
Dense DDFs can overfit locally in homogeneous tissue. On Dataset 1 (prostate), VoxelMorph and TransMorph achieved reasonable Dice but large landmark TRE and visibly distorted gland interiors (Fig. \ref{fig:prostate_visual}). Increasing the bending-energy weight reduced local distortion but also impeded convergence (Fig. \ref{fig:auto} (1)), reflecting a known smoothness–fidelity trade-off: excessive regularisation suppresses deformations, whereas weak regularisation permits implausible warps~\cite{elastixManual520, wang2019review}. Dice alone did not fully expose these issues, underscoring the need to consider geometric plausibility alongside overlap. 

\textcolor{revised}{
Prostate MR images often exhibit relatively homogeneous intensity with the prostate gland and substantial anatomical variation across patients. These characteristics make it challenging for deep networks to reliably identify consistent keypoints or correspondences between different subjects.  Constraining the degrees of freedom at prediction time using a simple coarse grid reduces the hypothesis space of admissible flows (often unnecessary local deformation), limiting spurious local distortions without sacrificing mask overlap. In contrast, brain MR (Dataset 3) provides abundant, consistent edges; all methods produced visually plausible deformations with similar centroid distances. }

\textcolor{revised}{
In prostate registration, the best-performing grids were over $20\times$ coarser than voxel-level DDFs, providing no evidence that a full-resolution parameterisation is necessary for these datasets. Instead, coarse grids benefit from inherent spatial regularisation, and performance was relatively insensitive to the exact grid size within a reasonable (albeit potentially application-dependent) range, from $5\times5\times5$ to $15\times15\times15$. We observed that using a grid size of approximately $10\%$ of the image size, in each dimension, led to generally good registration performance, based on Dataset~1.}

\textcolor{revised}{
Although Bayesian head did not improve registration performance, we included this uncertainty estimation for reference purposes for those interested in Figure~\ref{fig:uncertainty} and Table~\ref{tab:ablation}. These visualisations highlight two main observations: regions with weak local features (e.g. homogeneous areas such as ventricles) are assigned high uncertainty due to the lack of distinctive gradients for unambiguous alignment, whereas contours and high-contrast structures exhibit lower uncertainty. 
}
\paragraph{{Clinical relevance}}
Higher DSCs in the prostate and pelvis regions suggest that \textcolor{revised}{
GridReg could support future clinical applications in organ tracking and atlas construction, pending prospective validation and task-specific accuracy benchmarks.} The low memory consumption of GridReg makes it suitable for use in resource-constrained environments, such as mobile devices and edge computing systems. 

\section{Conclusion}
This work provides a systematic comparison of learning-based registration parameterisations that predict deformation from either (1) sparse, coarse regular grids of control points, (2) scattered non-gridded control points and (3) dense displacement fields. We show that regular grids offer a practical advantage: they enable fast and stable reconstruction of dense deformation fields via standard interpolation (e.g., trilinear) or transpose-convolution-based spline approximations, avoiding the additional complexity and overhead of resampling from irregular point sets. Importantly, a sparse deformation parameterisation provides implicit capacity control: fewer degrees of freedom suppress high-frequency voxel-wise warps driven by noisy or ambiguous correspondences, while reducing memory and compute.
\textcolor{revised}{GridReg further supports grid-adaptive training, allowing a single model to operate across multiple grid resolutions and to select an effective grid density for each dataset through validation, rather than committing to a fixed resolution a priori.} Finally, the approach is encoder-agnostic: it can be attached to different backbone encoders from existing registration networks with minimal changes. Overall, GridReg provides a simple, general, and efficient path to sparse deformation modelling, improving the efficiency–accuracy trade-off and broadening applicability across diverse medical image registration tasks.


\section*{Acknowledgment}
This work is supported by the International Alliance for Cancer Early Detection, an alliance between Cancer Research UK [C28070/A30912; C73666/A31378; EDDAMC-2021/100011], Canary Center at Stanford University, the University of Cambridge, OHSU Knight Cancer Institute, University College London and the University of Manchester. This work is also supported by the National Institute for Health Research (NIHR) University College London Hospitals (UCLH) Biomedical Research Centre (BRC).

%

\section*{Conflict of Interest Statement}
Mark Emberton receives research support from the United Kingdom’s National Institute of Health Research
(NIHR) UCLH/UCL Biomedical Research Centre. He acts as consultant/lecturer/trainer to Sonacare Inc.
Angiodynamics Inc. Early Health Ltd and Albemarle Medical Ltd The remaining authors declare that they
have no known competing financial interests or personal relationships that could have appeared to influence
the work reported in this paper.

\section*{Data availability}
Dataset 1 and Dataset 2 are retrospective datasets, which are not publicly available. Part of Dataset 2 is available for academic use upon formal application at \url{https://ncita.org.uk/promis-data-set-open-access-request}. 
Dataset 3 is publicly available at \url{https://learn2reg.grand-challenge.org}


\bibliography{bibliography}
\bibliographystyle{medphy.bst} 
\end{document}